\documentstyle[aps,titlepage]{revtex}

\newcommand{\be}{\begin{equation}}
\newcommand{\ee}{\end{equation}}
\newcommand{\ba}{\begin{eqnarray}}
\newcommand{\ea}{\end{eqnarray}}

\def\e{{\rm e}}

\begin{document}
\title{Dielectric function of a two-component plasma including collisions}

\author{G. R\"opke and  A. Wierling}
\address{FB Physik, Universit\"at Rostock, D-18051 Rostock, Germany}
\date{\today}

\maketitle

\begin{abstract}
A multiple-moment approach to the dielectric function of a dense non-ideal plasma
is treated beyond RPA 
including collisions in Born approximation. The results are compared
with the perturbation expansion of the Kubo formula. Sum rules
as well as Ward identities are considered. The relations to optical
properties as well as to the dc electrical conductivity are pointed out.
\vspace{1mm}
\\
\end{abstract}


\section{Introduction}

The dielectric function $\epsilon (\vec k, \omega)$ is a physical
quantity containing a lot of information about the plasma. In homogeneous,
isotropic systems it is
related to the electrical conductivity $\sigma(k, \omega)$ and the
polarisation function $ \Pi (k,\omega) $ according to
\be
\label{1}
\epsilon ( k, \omega) = 1+ {i \over \epsilon_0 \omega}\,\, \sigma
(k, \omega) = 1 - {1 \over \epsilon_0 k^2}\,\, \Pi (k,
\omega).
\ee
A well
established expression is the random phase approximation (RPA) 
valid for collisionless plasmas. The inclusion of collisions, however, 
is connected with difficulties. A perturbative treatment of the Kubo formula 
is not applicable near $\vec k = 0,\,\,\omega = 0$ because there is an essential 
singularity in zeroth order. Partial summations are sometimes in conflict 
with sum rules. Improvements of the RPA result are discussed in the static limit, 
where the local field corrections are treated in time-dependent mean-field theory 
\cite{GK}. Also approximations based on the sum rules for the lowest moments 
have been proposed \cite{A}. However, an unambigous expression for 
$\epsilon(\vec k, \omega)$ in the entire $\vec k\,\,
\omega$ space cannot be given by these approaches.

A particular problem is the appropriate treatment of the long-wavelength
limit $k \rightarrow 0$ at small frequencies where the dc conductivity
should be obtained. In a previous paper \cite{1} an  approach has been given 
where this limiting case coincides with the Chapman-Enskog approach
\cite{ChapCow} to the dc conductivity.
In particular, the polarization function was found as
\be
\label{2}
\Pi(k, \omega)= i\, {k^2 \over \omega}\,\,\beta\, \Omega_0
\left| \begin{array}{rr}   0 & M_{0n} (k,\omega)  \\
  M_{m0}(k,\omega)  & M_{mn}(k,\omega)
\end{array} \right| \Big/\, |M_{mn} (k,\omega) |.
\ee
The matrix elements $M_{mn}$ are equilibrium correlation function
which are explicitly given in the following section. They 
contain operators $B_m$ and $B_n$ which specify the nonequilibrium
state.

For the evaluation of the dielectric function, we have to deal with two problems:

i) the choice of the operators $B_n$ to describe the relevant
fluctuations in the linear response regime,

ii) the evaluation of the equilibrium correlation functions.

The equilibrium correlation functions in a nonideal plasma can be 
evaluated using the method of thermodynamic Green functions.
In lowest order of the perturbation theory to be considered here
we have the Born approximation as described in \cite{1}. Higher order 
terms can be taken into account in a systematic way, see \cite{R}.

With respect to the choice of the operators $B_n$, only the current density
operator $J$ has been considered in \cite{1}. In the spirit of the
Chapman-Enskog approach we will include here higher moments of
the single-particle distribution function to study the convergency
behavior. For the dc conductivity the answer is well known see \cite{RR}.
Note that different approaches based on
different sets of relevant observables $B_n$ are formally equivalent
as long as no approximations in evaluating the correlation functions
are performed. However, within a finite order perturbation theory,
the results for the conductivity are improved if the set of relevant
observables is extended.

Results for the dielectric function within a four-moment approach are
shown in Sec. II and compared with the results of a single-moment approach.
Some exact relations are discussed in Sec. III. Of particular interest 
is the relation to the Kubo formula which may be treated in perturbation 
theory as discussed in Sec. IV.

\section{Four-moment approach for a two-component plasma}

To evaluate the dielectric function we use the expression (\ref{2}) for the
polarisation function, where the matrix elements are given by
\ba
\label{3}
M_{0n}(k,\omega)& =& (J_{k};B_{n})\,\,,\qquad
M_{m0}(k,\omega) = (B_{m}; \hat{J}_{k})\,\,, \nonumber \\
M_{mn}(k,\omega)& =& (B_{m}; [\dot B_{n}-i \omega B_{n}]) +
\langle \dot B_{m};
[\dot B_{n}-i \omega B_{n}]\rangle_{\omega+i \eta}
- \frac{\langle \dot B_{m};
{J}_{k}\rangle_{\omega+i \eta}}{\langle B_{m};
{J}_{k}\rangle_{\omega+i \eta}}\,\, \langle B_{m};[\dot B_{n}-
i \omega B_{n}] \rangle_{\omega+i \eta}\,\,.
\ea

The equilibrium correlation functions are defined as
\ba
\label{4}
(A;B)&=& (B^+;A^+)={1 \over \beta} \int_0^\beta d\tau\,\, {\rm Tr} \left[
A(-i \hbar \tau) B^+ \rho_0 \right]\,\,, \nonumber\\
\langle A;B \rangle_z &=& \int^{\infty}_0 dt\,\, \e^{izt}\, (A(t);B)\,\,,
\ea
with $A(t)=\exp(iHt/\hbar)\,A\,\exp(-iHt/\hbar)$ and $\dot A ={i \over
  \hbar} [H,A]$ .
$\rho_0  = \exp (-\beta H + \beta \sum_c\mu_cN_c)\,\Big/\,{\rm Tr}
\exp (-\beta H + \beta \sum_c\mu_c N_c)$ is the equilibrium statistical 
operator.

We will consider a two-component plasma 
consisting of electrons ($c=e$) and ions ($c=i$). In particular, results 
are given below for a hydrogen plasma. With the single-particle operators 
\be
\label{5}
n^c_{p,k}= \left(n^c_{p,-k}\right)^+=c^+_{p-k/2}\,\,c^{}_{p+k/2}
\ee
the current density operator is given by
\be
\label{6}
J_k={1 \over \Omega_0}\,\, \sum_{c,p} {e_c
  \over m_c}\,\, \hbar p_z\,\, n_{p,k}^c\,\,.
\ee
Furthermore we used the abbreviation
$\hat{J}_{k} =  \epsilon^{-1}(k, \omega)\,\,J_k\,\,.$

To select the relevant operators $B_n$, we restrict us to the ordinary
kinetic approach. The inclusion of higher order correlations is also
possible, see \cite{R}.

Within the kinetic approach, the nonequilibrium state of the plasma is described
by the mean values of th single-particle operators (\ref{5}) corresponding to an induced
single-particle distribution function with wave number $k$.
Instead of treating an infinit number of operators depending on the
momentum $p$, we can restrict us to a finite number of moments of the
distribution function. This procedure is familiar from the theory of
the dc conductivity. Whereas in that case only moments with respect
to $p$ have to be selected, in the general case of arbitrary $k$
to be considered here moments of $p$ as well as $\vec p \cdot \vec k$
have to be taken into account.

In this paper we investigate how the lowest moment approach in Born 
approximation is modified if further moments are included. From the 
theory of dc conductivity we know that important modifications are
obtained by including the energy current density in addition to the particle
current density, i. e. if we include also $\vec p^2 p_z$. Then, the electrical 
cunductivity is not only described by the electron-ion interaction, but
includes also the effects of electron-electron interaction which are
not effective in the lowest moment approximation due to the conservation of total
momentum.

The four-moment approach to be considered in this paper is given
by the following moments of the electron ($c=e$) or ion ($c=i$) distribution
function, respectively,
\ba
\label{7}
&&b^c_1(p)={\hbar \over \sqrt{2m_c kT}} p_z,\nonumber\\
&&b^c_2(p)=\left({\hbar \over \sqrt{2m_c kT}}\right)^{3/2} (\vec{p})^2
p_z\,\,.
\ea
The evaluation of the corresponding correlation functions in Born 
approximation is given in the Appendix for the nondegenerate case. 
As a trivial result,
in the lowest approximation with respect to the interaction the
RPA result is recovered. In general the matrix elements are given in
terms of integrals of expressions containing the Dawson integral.

To give an example, a hydrogen plasma is considered with parameter 
values $T=98\,\, \mbox{Ryd}$ and $n_e=8.9\,\, a_B^{-3}$
 which are found in the center of the sun
\cite{sonne}. The results are comparable to the results obtained in
\cite{1} for parameter values corresponding to laser produced
high-density plasmas \cite{S}.  

Results for the real and the imaginary part of the dielectric function
in the two-moment approximation given by $b_1^c(p)$
are shown in figures 1 and 2, respectively. Besides the RPA dielectric function
the one-moment calculation reported in \cite{1} is shown as well. 
While the differences between the
improved dielectric function and the RPA are small at high momenta
($ k=1 \,\,a_B^{-1}$ ),  
significant changes occur at small momenta ($k=0.1\,\, a_B^{-1}$). 
On the other hand, the one-moment approach is almost identical
with the two-moment calculation. This is an indication that
convergence is reached by augmenting the number of moments as is
expected from earlier studies of the dc conductivity \cite{RR}. 
Note, that the static limit is given by the Debye law.

Results for the inverse dielectric function, which describes the
response to the external potential, are shown in figures 3 ($k=0.5
\,\,a_B^{-1}$) and 4 ($k=0.3 \,\,a_B^{-1}$) and compared with the RPA
inverse dielectric function.  Major deviations occur only at
frequencies close to the plasma frequency.  For small momenta, the
imaginary part of the dielectric function including collisions is
considerably broader compared with the RPA one. While the imaginary
part of the inverse dielectric function in the RPA approximation
becomes delta-like in the long wavelength limit, a broadening of the
plasmon peak appears, as can be seen from figure 4. Some properties of
the dielectric function will be discussed in the following section.

\section{Exact relations for the dielectric function and limiting cases}

Several exact properties of the dielectric function are known 
\cite{mahan} such as sum rules
\begin{eqnarray}
\label{fsum}
-\int_{-\infty}^{\infty}\!\frac{d\omega}{\pi}\,
 \omega\,\mbox{Im}\,\epsilon^{-1}(k,\omega)
 & = & \omega_{pl}^2\;\;\;, \\
\label{conduct}
\int_{-\infty}^{\infty}\!\frac{d\omega}{\pi}\,
 \omega\,\mbox{Im}\,\epsilon(k,\omega) & = & \omega_{pl}^2 \;\;\;,\\
\label{total}
  \lim_{k\to 0}
  \int_{-\infty}^{\infty}\!\frac{d\omega}{\pi}\,\frac{1}{\omega}\,
\mbox{Im}\,\epsilon^{-1}(k,\omega) & = & -1\;\;\;,
\end{eqnarray}
the long-wavelength limit
\begin{eqnarray}
  \label{compress}
  \lim_{k\to 0} \mbox{Re}\,\epsilon(k,0) & = & 1+V(k)\,n^2\,K\;\;\;.
\end{eqnarray}
Here $\omega_{pl}^2\;=\;
\sum_{c=e,i} \left(e^2\,n_c\right)/\left(\epsilon_0\,m_c\right)$ denotes the
plasma frequency and $K$ the isothermal compressibility. Further extensions for
 a two-component system can be found in \cite{rasolt}.
This is a special relation resulting from the relation between 
the dynamical structure factor 
\begin{eqnarray}
  S(k,\omega) & = & \frac{1}{2\,\pi}\,\int_{-\infty}^{\infty}
  dt\,<\rho_k^+(t)\,\rho_k>\,e^{i\omega t}
\end{eqnarray} and the dielectric function which can be established via the
fluctuation-dissipation-theorem
\begin{eqnarray}
  S(k,\omega) & = & -\frac{1}{\pi}\,\frac{1}{e^{\beta\,\omega}-1}\,
  \mbox{Im}\,\epsilon^{-1}(k,\omega^-)\;\;\;.
\end{eqnarray}
Furthermore, the Kramers-Kronig relation holds which connects the real
and the imaginary part of the dielectric function:
\begin{eqnarray}
\mbox{Re} \epsilon(k,\omega) & = &
1\,+\,P\int\!\frac{d\omega'}{\pi}\,
\frac{\mbox{Im}\,\epsilon(\vec k,\omega')}{\omega-\omega'}\,\,.
\end{eqnarray}
Here, $P$ denotes the Cauchy principal value integration. 
The inverse dielectric function obeys a corresponding relation.
Combining the Kramers-Kronig relation with the sum rules results
in rigorous statements about the asymptotic behaviour at high frequencies:
\begin{eqnarray}
  \lim_{\omega \to \infty} \mbox{Re} \epsilon(k,\omega) & = &
  1-\frac{\omega_{pl}^2}{\omega^2}+O(\frac{1}{\omega^4})\;\;\;.
\end{eqnarray}

We test the two-moment approach by checking the sum rules as well as 
the asymptotic behaviour. It is found that the sum rules are fulfilled
within the numerical accuracy ($ \approx 0.1 \% $). The Kramers-Kronig
relation holds as well. 
Having in mind relation (\ref{1}), the dc conductivity can be obtained
considering the limitng case $k \rightarrow 0$. A comparison 
with other results for the conductivity can be made by parameterising the
conductivity via 
\begin{eqnarray}
  \sigma=\sigma(0,0) & = & s
  \frac{(k_B\,T)^{3/2}\,(4\pi\epsilon_0)^2}{e^2\,m_{ei}^{1/2}}\,
  \frac{1}{\Phi}\;\;\;,
\end{eqnarray}
where $\Phi$ denotes the Coulomb-logarithm and $m_{ei}$ the reduced mass
of the electron. 



As shown in figure 5, there is no shift of the maximum of
the inverse dielectric function, while the plasmon peak is broadened.
Moreover, the long wavelength limit can be described by a Drude-like
formula, implying that the width of the plasmon peak is given by the
dc conductivity. The form of the plasmon peak can be compared with
computer simulation studies. In contrast to RPA calculations, width as
well as height of the plasmon peak in our calculation
are of comparable size as computer simulations \cite{Ort}.

\section{Comparison with the Kubo formula}

Depending on the selected set of relevant operators $\{B_n\}$,
different expressions for the dielectric function can be
derived within linear response theory. A often used expression
is the Kubo formula \cite{Z} as given by 
\be
\label{20}
\Pi(k, \omega) = -\,\,{i k^2 \beta \Omega_0 \over \omega}\,\,
\langle J_{k};\hat J_{k} \rangle_{\omega +i \eta}\,\,.
\ee
As shown in \cite{1}, this result follows as a special case
within the generalized linear response theory. As also shown there,
the different expressions identical in the limit $\eta \rightarrow 0$
if no further approximations are performed.

The advantage of linear response theory is that the evaluation of
the dielectric function is related to the evaluation of equilibrium
correlation functions. In dense, strongly coupled systems, these
correlation functions can be calculated with computer simulations.
Another possibility is to use peturbation theory which is most
effectively formulated with the concept of thermodynamic Green
functions \cite{KKER}.

In zeroth order with respect to the interaction, from (\ref{20})
immediately the RPA result is obtained, in coincidence with all
other approaches including $J$ within the set of relevant operators. 
The first order expansion with respect to the screened interaction reads
\ba
&& \Pi (k, \omega_\lambda) = \sum_p \left(f_p+f'_p n_{\rm ion} \sum_q V_q^2
{1 \over E_{p}-E_{p-q}} \right) \left({1 \over E_p - \omega_\lambda
  -E_{p-k}} + {1 \over E_p + \omega_\lambda -E_{p+k}} \right)
\nonumber\\
&& + n_{\rm ion} \sum_{p q} V_q^2 f_p {kq \over m} {1 \over E_p - E_{p-q}} 
{1 \over E_p -\omega_\lambda-E_{p-k}}{1 \over E_p - \omega_\lambda -E_{p-k-q}} 
\left(  {1 \over E_p -E_{p-q}}+
{1 \over E_p - \omega_\lambda-E_{p-k}} \right)\nonumber\\
&& +(\omega,k \leftrightarrow - \omega, -k).
\ea
For the sake of simplicity, we have taken the adiabatic limit where
$m_i/m_e \rightarrow \infty$ (Lorentz plasma),
In particular we find for $k \rightarrow 0$
\be
{\rm Im}\Pi (k, \omega) =  n \sum_{pq} V_q^2 \left({kq \over
  m}\right)^2 \pi \delta(E_p-\omega-E_{p-q}) \e^{-\beta (E_p-\mu)} {1
  - \e^{\beta \omega} \over \omega^4}.
\ee
what gives the frequency-dependent conductivity.

However, this perturbation expansion does not converge at $\omega 
\rightarrow 0$, and partial summations have to be performed.
For instance, a simple approximation for the polarization function 
including interactions with further 
particles would be a polarization function given by the product 
of two full propagators. This way, the polarization
function contains shifts and damping of the single-particle states due
to the interaction with the medium.
However, this approximation does not fulfill rigorous relations such
as sum rules, since important corrections to the RPA of the 
same order in the density as the considered ones are missing,
e.g. vertex corrections. These corrections are linked to the
self-energy by Ward identities \cite{WT}. As a consequence, the vertex
has to be improved in accordance with the self-energy. 
Following Baym and Kadanoff \cite{BaymKad}, a consistent vertex can be constructed to
a given self-energy. 
However, the solution of the vertex equation cannot be given in a
simple algebraic form, and usually some approximations are performed,
see \cite{wir}.

\section{Conclusions}

An approach to the dielectric function  has been investigated which
includes the effects of collissions and can be used in the entire
$k,\omega$ space. Within a four-moment approach to a two-component
plasma, the Born approximation has been evaluated, and important
rigorous results for the dielectric function are checked. Compared
with the ordinary Kubo formula, the approach given here seems to
be more appropriate for perturbation expansions.

In particular, comparing with a one-moment approach, the convergency
behavior of this method was inspected. As well known from the
theory of dc conductivity, convergence is expected if higher moments are 
included. In a more general approch, also two-particle correlations
can be included into the set of relevant operators.

Within a quantum statistical approach, the Born approximation can be
improved by systematic treatment of Green functions. This concerns,
e.g., the inclusion of strong collision by treating T-matrices, 
degeneracy effects, and the treatment of the dynamic screening of 
the interaction. Here, the comparison with computer simulations is
also an interesting perspective. Work in this direction is in progress.

\vspace{1cm}

The authors acknowledge helpful discussions with W. Ebeling, W.D. Kraeft, 
D. Kremp, R. Redmer, and Chr. Toepffer. 

\section*{Appendix: Evaluation of the matrix elements of $\Pi$}

We start from the general expression (1), (2) for the dielectric function
with ($n,m$ =1...4,$\qquad c,d=e,i$)
\be
M^d_{0m}(q,\omega)={1 \over \Omega_0 } \sum_{c,p,k} {e_c \over m_c} {q
  \over \omega} \hbar p_z b^d_m(k) (n^d_{k,-q};n^c_{p,q})\,\,,
\ee
\be
M^c_{n0}(q,\omega)={1 \over \Omega_0 } \sum_{d,p,k} iq {e_d \over m_d}
 \hbar k_z [b^c_n(p)]^* (n^d_{k,-q};n^c_{p,q})\,\,,
\ee
\be
M^{cd}_{nm}(q,\omega)={1 \over \Omega_0 } \sum_{p,k} [b^c_n(p)]^* 
b^d_m(k) ([\dot{n}^d_{k,-q}-i\omega n^d_{k,-q}];n^c_{p,q})+A
\ee
or, after some rearrangements,
\be
\epsilon (q,\omega)=1- {\beta n e^2 \over \epsilon_0 q \omega}\left| 
{0 \qquad \tilde{M}_{0m}^d (q,\omega) \over   \tilde{M}_{n0}^c
  (q,\omega) \qquad \tilde{M}_{nm}^{cd} 
  (q,\omega)} \right|/ |\tilde{M}_{nm}^{cd} (q,\omega) |
\ee
with
\be
\tilde{M}_{0m}^d (q,\omega)={z_d \over ne} M_{0m}^d (q,\omega)
={z_d \over ne}{1 \over \Omega_0 } \sum_{c,p,k} {e_c \over m_c} {q
  \over \omega} \hbar p_z b^d_m(k) (n^d_{k,-q};n^c_{p,q})\,\,,
\ee
\be
\tilde{M}_{n0}^c (q,\omega)=-i {1 \over \omega}{z_c \over ne} 
M_{n0}^c (q,\omega)=-i {1 \over \omega}{z_c \over ne}
{1 \over \Omega_0 } \sum_{d,p,k} iq {e_d \over m_d}
 \hbar k_z [b^c_n(p)]^* (n^d_{k,-q};n^c_{p,q}) \,\,,
\ee
\ba
&&\tilde{M}_{nm}^{cd} (q,\omega)=-i {\sqrt{m_cm_d}\over 2 kT nq}
M_{nm}^{cd} (q,\omega)=-i {\sqrt{m_cm_d}\over 2 kT nq}
{1 \over \Omega_0 } \sum_{p,k} [b^c_n(p)]^* 
b^d_m(k) \left\{ ([\dot{n}^d_{k,-q}-i\omega n^d_{k,-q}];n^c_{p,q}) \right.\nonumber\\
&&+ \left. \langle \dot {n}^d_{k,-q};
[\dot {n}^c_{p,q}-i \omega n^c_{p,q}]\rangle_{\omega+i \eta}
- \frac{\langle \dot {n}^d_{k,-q};
{J}_{k}\rangle_{\omega+i \eta}}{\langle {n}^d_{k,-q};
{J}_{k}\rangle_{\omega+i \eta}}\,\, \langle {n}^d_{k,-q};[\dot n^c_{p,q}-
i \omega n^c_{p,q}] \rangle_{\omega+i \eta} \right\}\,\,,
\ea
and
\be
z_c={\omega \over q} \sqrt{{m_c \over 2kT}}\,\,.
\ee

We specify to a four-moment approach (\ref{7}) where $B_1=b^e_1(p),\,\,
B_2=b^e_2(p),\,\,B_3=b^i_1(p),\,\,B_4=b^i_1(p)$.
Introducing the Dawson integral
\be
D(z)= \lim_{\delta \rightarrow +0} {1\over \sqrt{\pi}}
  \int_{-\infty}^\infty dx {\rm e}^{-x^2} {1 \over x-z-i\delta} 
\ee
and using the abbreviations
\be
r^c_1={1\over 2}\,\,{1 \over 1+z_c D(z_c)}\,\,,
\ee
\be
r^c_2={5\over 4}\,\,{1 \over 0.5+(1+z_c^2) [1+z_c D(z_c)]}\,\,,
\ee
we have for $n_e=n_i=n$, $e_e=-e_i=e$
\be
\tilde{M}^e_{01}={1 \over 2}, \qquad
\tilde{M}^e_{02}={5 \over 4}, \qquad
\tilde{M}^i_{03}=-{1 \over 2}, \qquad
\tilde{M}^i_{04}=-{5 \over 4}, \qquad
\ee
and
\be
\tilde{M}^e_{10}={1 \over 2}, \qquad
\tilde{M}^e_{20}={5 \over 4}, \qquad
\tilde{M}^i_{30}=-{1 \over 2}, \qquad
\tilde{M}^i_{40}=-{5 \over 4}, \qquad
\ee
We decompose
\be
\tilde{M}_{nm}^{cd} (q,\omega)=a_{nm}+b_{nm}+c_{nm}
\ee
and find in zeroth order with respect to the interaction
\be
a_{11}={1\over 2} r^e_1 {q \over \omega},\qquad
a_{12}={5\over 4} r^e_1 {q \over \omega},\qquad
a_{21}={5\over 4} r^e_2 {q \over \omega},\qquad
a_{22}={35\over 8} r^e_2 {q \over \omega},
\ee
and
\be
a_{33}={1\over 2} r^i_1 {q \over \omega},\qquad
a_{34}={5\over 4} r^i_1 {q \over \omega},\qquad
a_{43}={5\over 4} r^i_2 {q \over \omega},\qquad
a_{44}={35\over 8} r^i_2 {q \over \omega}.
\ee

The $b_{nm}$ contain the electron-ion interaction in first Born
approximation and the $c_{nm}$ the electron-electron or ion-ion
interaction, respectively. We use a screened interaction with the
Debye screening factor $\exp (- \kappa r),\,\,\kappa^2=\sum_c n_c
e_c^2 /(\epsilon_0 kT)$.

Terms due to electron-ion interaction are with $M=m_e+m_i$
\be
b_{ij}= -{i \over 8 (2 \pi)^{3/2} } {1 \over q} n {e^4 \over
  \epsilon_0^2} \left( {1 \over kT} \right)^{5/2} \left( {m_e m_i \over
    M} \right)^{1/2} g_{ij}\,\,.
\ee
With
\be
z_{ei}={\omega \over q} \sqrt{{M \over 2kT}}
\ee
and 
\be
\lambda^{ei}=1+{\hbar^2 \kappa^2 M \over 4 m_e m_i kT}\,{1 \over p^2}\,\,,
\ee

\be
\Lambda_1= [ \ln \left(
{\lambda^{ei}-1 \over \lambda^{ei}+1} \right) +{2 \over
  \lambda^{ei}+1}]\,\,,
\ee
\be
\Lambda_2= [\lambda^{ei} \ln \left(
{\lambda^{ei}-1 \over \lambda^{ei}+1} \right) +2 ],\qquad
\Lambda_3={2 \over
  (\lambda^{ei})^2-1}\,\,,
\ee
\be
R^c_n ={M\over \sqrt{m_e m_i}} r^c_n\,\,,
\ee
\be
D_e= D( z_{ei}- \sqrt{{m_i \over m_e}} pc),\qquad
D_i= D( z_{ei}- \sqrt{{m_e \over m_i}} pc)\,\,,
\ee
we find
\be
g_{11}  =  \int_0^\infty dp {\rm e}^{-p^2}\,\Lambda_1 \left\{ {2 \over3} p
 - R^e_1 \int_{-1}^1 dc \, c (2 D_e+ D_i)\right\}\,\,,
\ee

\be
g_{13}  =   \int_0^\infty dp {\rm e}^{-p^2}\,\Lambda_1 \left\{ -{2 \over3}
p  + R^e_1 \int_{-1}^1 dc \, c D_e \right\}\,\,,
\ee

\ba
g_{12}&= & \int_0^\infty dp {\rm e}^{-p^2}\,\left\{ \Lambda_1 ({5
  \over 3} {m_e \over M} p+{2\over 3} {m_i \over M} p^3+R^e_1 2 
{ \sqrt{m_e m_i} \over M} p)+ \right. \nonumber\\
&+&\int_{-1}^1 dc R^e_1 (D_e \left[\Lambda_1 c (-{5 \over 2}- {m_e \over M}
- {m_i \over M} p^2 -3  {m_e \over M} (z_{ei}- \sqrt{{m_i \over m_e}}
pc)^2)+\Lambda_2 2 p{ \sqrt{m_e m_i} \over M} (1-3 c^2) (z_{ei}-
\sqrt{{m_i \over m_e}} 
pc) \right]\nonumber\\
& +&\left. D_i \Lambda_1 c (-{5 \over 2})) \right\}\,\,,
\ea
\ba
&&g_{14}=  \int_0^\infty dp {\rm e}^{-p^2}\,\left\{ \Lambda_1 (-{5
  \over 3} {m_i \over M} p-{2\over 3} {m_e \over M} p^3-R^e_1 2 
{ \sqrt{m_e m_i}m_i \over M m_e} p)+ \right. \nonumber\\
&+&\left. \int_{-1}^1 dc R^e_1 (D_e \left[\Lambda_1 c ( {m_i \over M}
+ {m_e \over M} p^2 +3  {m_i \over M} (z_{ei}- \sqrt{{m_i \over m_e}}
pc)^2)+\Lambda_2 2 p{ \sqrt{m_e m_i} \over M} (1-3 c^2) (z_{ei}-
\sqrt{{m_i \over m_e}} 
pc) \right]) \right\}\,\,,
\ea

\ba
&&g_{21}=  \int_0^\infty dp {\rm e}^{-p^2}\,\left\{ \Lambda_1 p({5
  \over 3} {m_e \over M} +{2\over 3} {m_i \over M} p^2+R^e_2 ({4
  \over 3} { \sqrt{m_e m_i} \over M} + 2 {m_e \sqrt{m_e m_i} \over
  m_i M}  )) +\int_{-1}^1 dc R^e_2 (D_e \right. \nonumber\\
&&\times \left[\Lambda_1 c (- 2 {m_e \over M}
- 2 {m_i \over M} p^2- 2 c { \sqrt{m_e m_i} \over M} p (z_{ei}-
\sqrt{{m_i \over m_e}} pc) - 4 {m_e \over M} (z_{ei}- \sqrt{{m_i \over m_e}}
pc)^2)+\Lambda_2 2 p{ \sqrt{m_e m_i} \over M} (1-3 c^2) (z_{ei}-
\sqrt{{m_i \over m_e}} pc) \right]\nonumber\\
& +&\left. D_i \left[ \Lambda_1 c (-{m_e \over M}
-  {m_i \over M} p^2 - 3 {m_e \over M} (z_{ei}- \sqrt{{m_e \over m_i}}
pc)^2)-\Lambda_2 2 p{ \sqrt{m_e m_i} \over M} (1-3 c^2) (z_{ei}-
\sqrt{{m_e \over m_i}} pc) \right]) \right\}\,\,,
\ea

\ba
g_{23}&= & \int_0^\infty dp {\rm e}^{-p^2}\,\left\{ \Lambda_1p \left(-{5
  \over 3} {m_e \over M} -{2\over 3} {m_i \over M} p^2+R^e_2 {2
  \over 3} { \sqrt{m_e m_i} \over M} \right) + \right. \nonumber\\
&+&\left. \int_{-1}^1 dc R^e_2 D_e \Lambda_1 c \left[ {m_e \over M}
+ {m_i \over M} p^2+ 2 c { \sqrt{m_e m_i} \over M} p (z_{ei}-
\sqrt{{m_i \over m_e}} pc) + {m_e \over M} (z_{ei}- \sqrt{{m_i \over m_e}}
pc)^2 \right] \right\}\,\,,
\ea

\ba
g_{22}&= & \int_0^\infty dp {\rm e}^{-p^2}\,\left\{ \Lambda_1 ({47
  \over 6} {m_e^2 \over M^2} p+{10\over 3} {m_e m_i \over M^2}
p^3+{2\over 3} {m_i^2 \over M^2} p^5\right. \nonumber\\
&&+R^e_2 { \sqrt{m_e m_i} \over M} (-{10
  \over 3}{m_e \over M} p+ 7 {m_e \over m_i} p+7 p+ {11 \over 3} {m_e \over
  M} p )+ 2 {m_e \over M} z^2_{ei} p+ {2 \over 15} {m_i \over M} p^3)
\nonumber\\ 
&&+ \Lambda_2 ({8 \over 15} {m_e^2 \over M^2} p+{40\over 15} {m_e m_i \over M^2}
p^3+R^e_2 { \sqrt{m_e m_i} \over M} (-{16
  \over 15}{m_e \over M} p+ {16 \over 15} {m_i \over M}
p^3))+\Lambda_3(-{4 \over 15} {m_e^2 \over M^2} p+R^e_2 { \sqrt{m_e m_i} \over M}{8
  \over 15}{m_e \over M} p ))\nonumber\\
&&+\int_{-1}^1 dc R^e_2 (D_e 
 \left[-\Lambda_1 c
({7 \over 2} {m_i \over M} p^2+{m_i^2 \over M^2} p^4+ 2 c{m_i \over M}
{ \sqrt{m_e m_i} \over M}p^3 (z_{ei}-\sqrt{{m_i \over m_e}}c p)
+ {7\over 2}{m_e \over M}(1+3 (z_{ei}-\sqrt{{m_i \over m_e}}c p)^2)\right.\nonumber\\
&&+2 {m_e m_i \over M^2}p^2(1+2(z_{ei}-\sqrt{{m_i \over m_e}}c p)^2)+
2 c {m_e \over M} { \sqrt{m_e m_i} \over M} p (z_{ei}-\sqrt{{m_i \over
    m_e}}c p) [1+3(z_{ei}-\sqrt{{m_i \over m_e}}c p)^2]\nonumber\\
&& + {m_e^2 \over
  M^2} [2+4(z_{ei}-\sqrt{{m_i \over m_e}}c p)^2+3 (z_{ei}-\sqrt{{m_i
    \over m_e}}c p)^4 ] ) + \Lambda_2 { \sqrt{m_e m_i} \over M}
p\nonumber\\
&&\times
(-6 c (1-c^2) p { \sqrt{m_e m_i} \over M}-2 (1-c^2)(1-3 c^2){m_e \over M} 
(z_{ei}-\sqrt{{m_i \over m_e}}c p) +(1-3 c^2)  (z_{ei}-\sqrt{{m_i
    \over m_e}}c p) [7 +2 {m_i \over M} p^2\nonumber\\
&& +4 c  p { \sqrt{m_e m_i}
  \over M}(z_{ei}-\sqrt{{m_i \over m_e}}c p) +2{m_e \over M} (1+ 
(z_{ei}-\sqrt{{m_i \over m_e}}c p)^2) ]   )\nonumber\\
&& \left. +2 \Lambda_3{ \sqrt{m_e m_i} \over M}{m_e \over M} p c^2 (1-c^2)
(z_{ei}-\sqrt{{m_i \over m_e}}c p)  \right] \nonumber\\
& +&\left. D_i \left[ \Lambda_1 c (-{7 \over 2} {m_e \over M} 
- {7 \over 2} {m_i \over M} p^2 - {21 \over 2} {m_e \over M} (z_{ei}- \sqrt{{m_e \over m_i}}
pc)^2)-\Lambda_2 7 p{ \sqrt{m_e m_i} \over M} (1-3 c^2) (z_{ei}-
\sqrt{{m_e \over m_i}} pc) \right]) \right\}\,\,,
\ea

\ba
g_{24}&= & \int_0^\infty dp {\rm e}^{-p^2}\,\left\{ \Lambda_1 p (-{47
  \over 6} {m_e m_i \over M^2} -{5\over 3} {m_e^2 \over M^2} p^2-
{5\over 3} {m_i^2 \over M^2}
p^2-{2\over 3} {m_e m_i \over M^2} p^4\right. \nonumber\\ 
&&+R^e_2 { \sqrt{m_e m_i} \over M} (-{1
  \over 3}{m_i \over M} - 2 {m_i \over M} z^2_{ei} + {2 \over 3} {m_e
  \over M} p^2-{4\over 5} {m_i^2\over m_e M}p^2)
\nonumber\\ 
&&+ \Lambda_2 p (-{8\over 15} {m_e m_i \over M^2}+{40 \over 15} {m_e m_i \over M^2}
p^2+{16 \over 15}{ \sqrt{m_e m_i} \over M}{m_i \over M} R^e_2(1+p^2))+\Lambda_3 p
({4 \over 15} {m_e m_i \over M^2}-R^e_2 { \sqrt{m_e m_i} \over M}{8 
  \over 15}{m_i \over M} )\nonumber\\
&&+\int_{-1}^1 dc R^e_2 (D_e 
 \left[\Lambda_1 c{ \sqrt{m_e m_i} \over M}
({ \sqrt{m_e m_i} \over M} p^4+ 2 c{m_e \over M}
p^3 (z_{ei}-\sqrt{{m_i \over m_e}}c p)
+\sqrt{{m_e \over m_i}} { m_e \over M}p^2(1+(z_{ei}-\sqrt{{m_i \over
    m_e}}c p)^2)\right.\nonumber\\ 
&&+\sqrt{{m_i \over m_e}} { m_i \over M}p^2(1+3(z_{ei}-\sqrt{{m_i \over m_e}}c p)^2)+
2 c  { m_i \over M} p (z_{ei}-\sqrt{{m_i \over
    m_e}}c p) [1+3(z_{ei}-\sqrt{{m_i \over m_e}}c p)^2]\nonumber\\
&& + { \sqrt{m_e m_i} \over M} [2+4(z_{ei}-\sqrt{{m_i \over m_e}}c p)^2+3 (z_{ei}-\sqrt{{m_i
    \over m_e}}c p)^4 ] ) \nonumber\\
&&+ \Lambda_2 { \sqrt{m_e m_i} \over M} [6c(-1+c^2)
{ \sqrt{m_e m_i} \over M} p^2+2 (1-c^2) (1-3 c^2){m_i \over M}p
(z_{ei}-\sqrt{{m_i \over m_e}}c p)\nonumber\\
&&+(1-3 c^2) p
(z_{ei}-\sqrt{{m_i \over m_e}}c p) (2{m_i \over M} p^2+4 c p
{ \sqrt{m_e m_i} \over M}
(z_{ei}-\sqrt{{m_i \over m_e}}c p)+2 {m_e \over M}
(1+(z_{ei}-\sqrt{{m_i \over m_e}}c p)^2))]\nonumber\\
&&-\left. \left.\Lambda_3 { \sqrt{m_e m_i} \over M}{m_i \over M}2 c^2
(1-c^2)p (z_{ei}-\sqrt{{m_i \over m_e}}c p) \right]) \right\}\,\,.
\ea
The remaining expressions  ($i =3,4$) follow as
\ba
&&g_{31}=[g_{13},(e \leftrightarrow i)],\qquad
g_{32}=[g_{14},(e \leftrightarrow i)],\qquad
g_{33}=[g_{11},(e \leftrightarrow i)],\qquad
g_{34}=[g_{12},(e \leftrightarrow i)],\nonumber\\
&&g_{41}=[g_{23},(e \leftrightarrow i)],\qquad
g_{42}=[g_{24},(e \leftrightarrow i)],\qquad
g_{43}=[g_{21},(e \leftrightarrow i)],\qquad
g_{44}=[g_{22},(e \leftrightarrow i)].
\ea

For the collisions between identical species $(e,i)$ we have
\be
c^c_{ij}= -{i \over 8 (2 \pi)^{3/2}} {1 \over q} n {e^4 \over
  \epsilon_0^2} \left({1 \over kT} \right)^{5/2} \left( {m_c \over
    2}\right)^{1/2} h_{ij}
\ee 
and
\be
\lambda^c=1+{\hbar^2 \kappa^2 \over 2 m_c kT}\,{1 \over p^2}
\ee

so that the contributions of electron-electron collisions $(i,j=1,2)$
follow as
\be
h_{11}=0\,\,,
\ee
\be
h_{12}=4 r^e_1 \int_0^\infty dp \int_{-1}^1 dc {\rm e}^{-p^2}\,p (1-3
c^2) [\lambda^e \ln \left( {\lambda^e-1 \over \lambda^e+1} \right) +2]
(\sqrt{2} z_e-c p ) D(\sqrt{2} z_e-c p )\,\,,
\ee
\be
h_{21}=4 r^e_2 \int_0^\infty dp \int_{-1}^1 dc {\rm e}^{-p^2}\,p (1-3
c^2) [\lambda^e \ln \left( {\lambda^e-1 \over \lambda^e+1} \right) +2]
(\sqrt{2} z_e-c p ) D(\sqrt{2} z_e-c p )\,\,,
\ee
\ba
&&h_{22}= \int_0^\infty dp {\rm e}^{-p^2}\, [\lambda^e \ln \left(
{\lambda^e-1 \over \lambda^e+1} \right) +2] \left\{{4 \over 3}p^3+{16
  \over 15}r^e_2p^3 +2 r^e_2  \int_{-1}^1 dc \,p \right. \nonumber\\
&&\times \left.[(1-3 c^2)(\sqrt{2} z_e-c p ) (p^2-p^2c^2 +8+2 z_e^2)+3 p
 c(c^2-1)] D(\sqrt{2} z_e-cp ) \right\}\,\,.
\ea
The expressions for ion-ion collisions ($i,j =3,4$) follow as
\be
h_{33}=[h_{11},(e \leftrightarrow i)],\qquad
h_{34}=[h_{12},(e \leftrightarrow i)],\qquad
h_{43}=[h_{21},(e \leftrightarrow i)],\qquad
h_{44}=[h_{22},(e \leftrightarrow i)],
\ee
i.e. replacing the index $e$ in $c^e_{ij},\,\,\lambda^e,\,\,z_e$ by
the index $i$.

\newpage
Figure Captions:

\vspace{3mm}

Fig. 1:

Real and imaginary part of the dielectric function as a function of
the frequency at fixed wavenumber $k=1\,\,a_B^{-1}$. The two-moment
approach is compared with the one-moment approach and the RPA. 

\vspace{3mm}

Fig. 2:

The same as Fig. 1 for wavenumber  $k=0.1\,\,a_B^{-1}$.

\vspace{3mm}

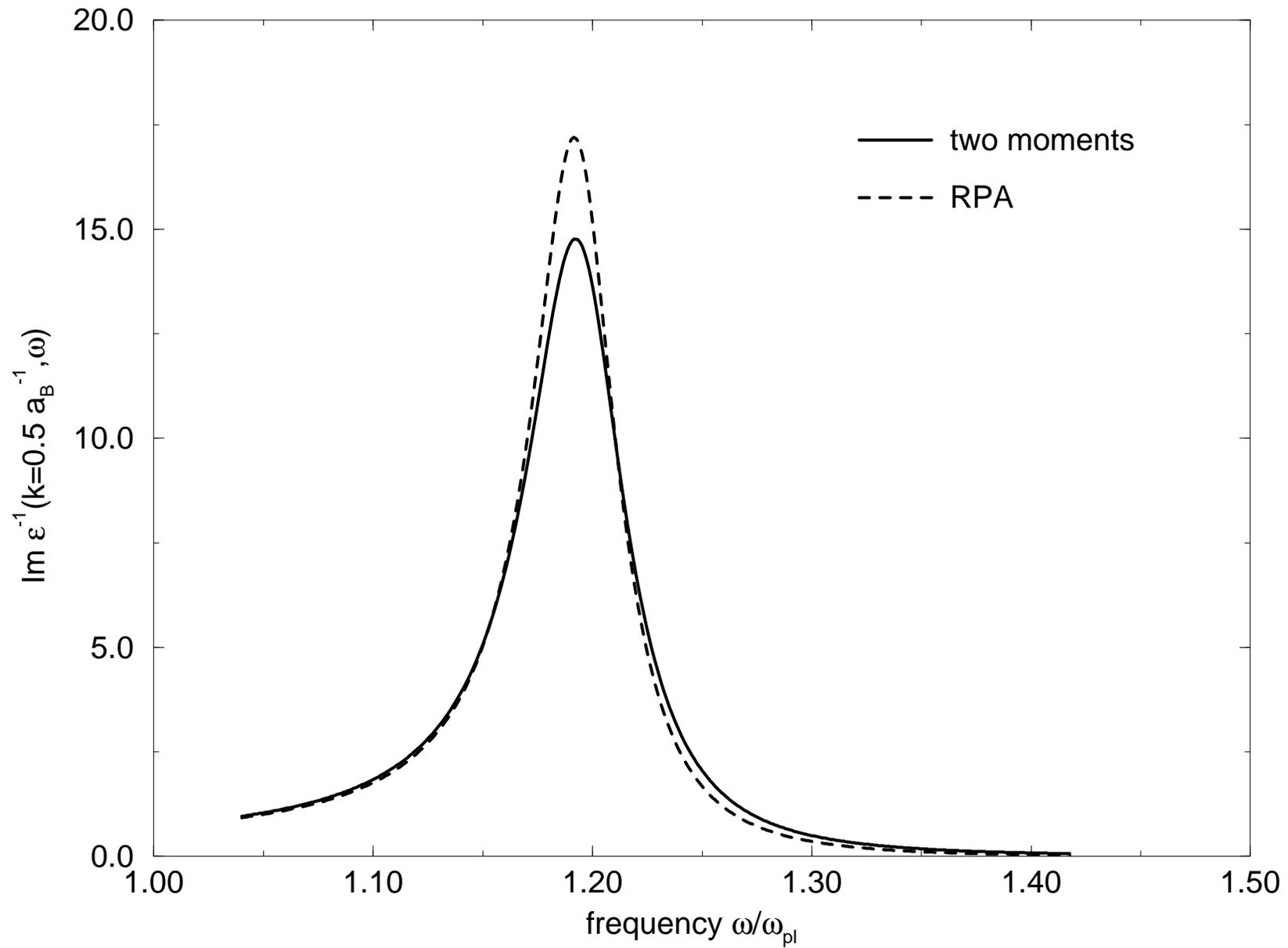
Fig. 3:

Imaginary part of the inverse dielectric function as a function of
the frequency at fixed wavenumber $k=0.5\,\,a_B^{-1}$. The two-moment
approach is compared with the RPA. 

\vspace{3mm}

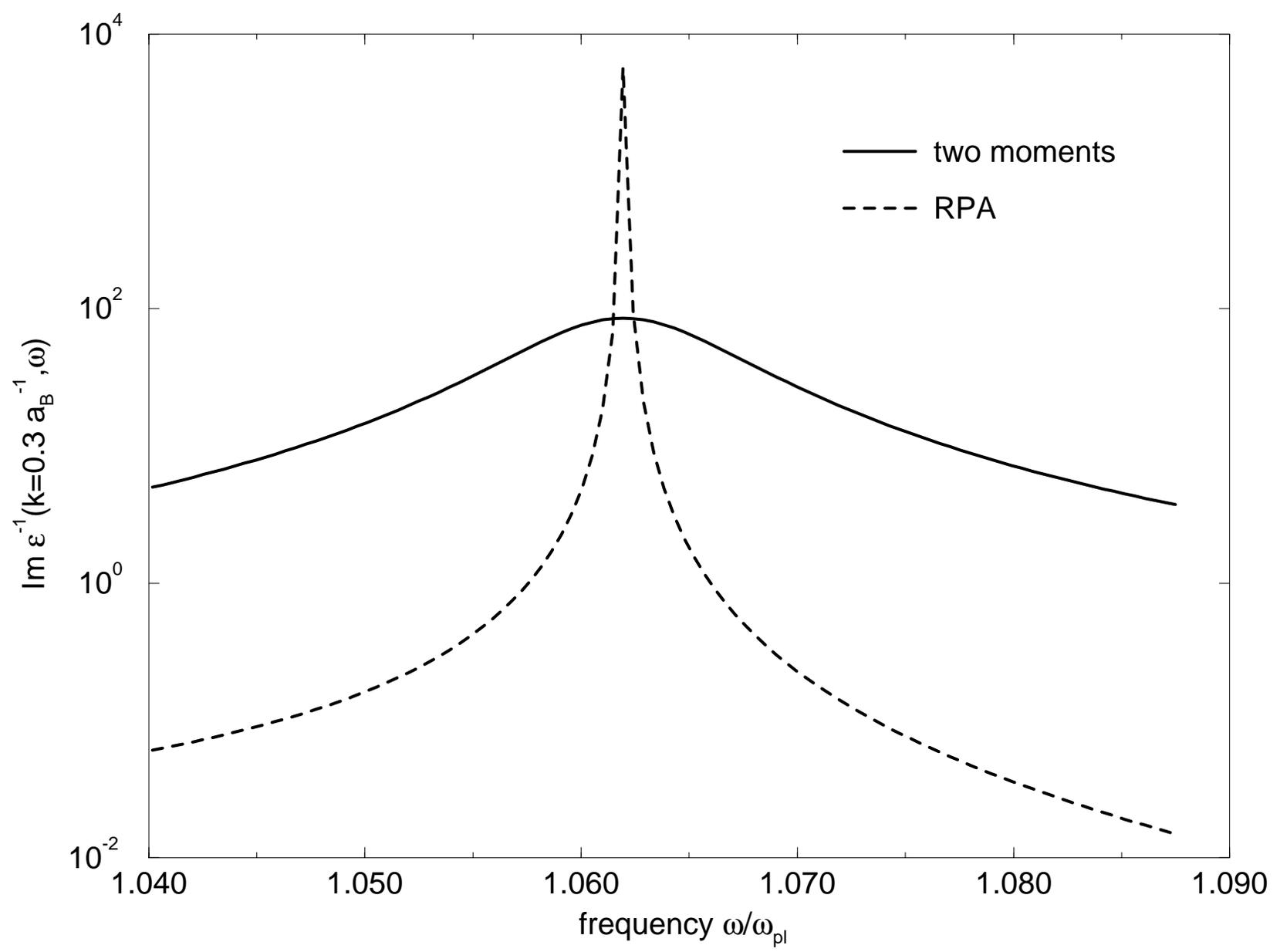
Fig. 4:

The same as Fig. 3 for wavenumber  $k=0.3\,\,a_B^{-1}$.

\vspace{3mm}

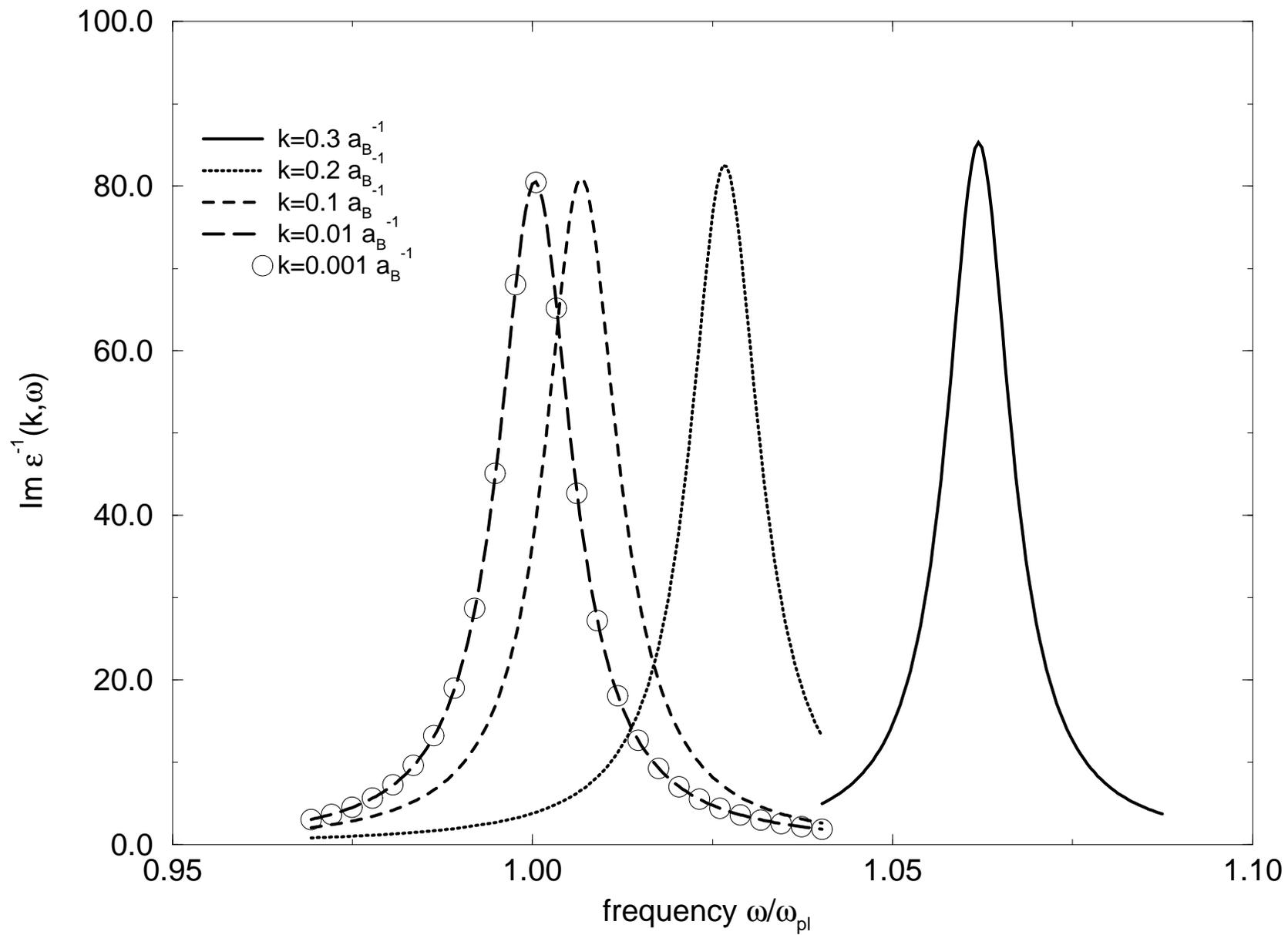
Fig. 5:

Imaginary part of the inverse dielectric function as a function of
the frequency at different wavenumbers.


\begin{thebibliography}{99}

\bibitem{GK}
K. I. Golden and G. Kalman, Phys. Rev. {\bf 19}, 2112 (1979).

\bibitem{A}
V. M. Adamyan and I. M. Tkachenko, Teplofiz. Vys. Temp. {\bf 21},
417 (1983) [Sov. Phys. High Temp. Phys. {\bf 21}, 307 (1983)];
J. Hong and M. H. Lee, Phys. Rev. Lett. {\bf 70}, 1972 (1993).

\bibitem{1}
G. R\"opke, preprint [physics/9709018]

\bibitem{ChapCow}
S. Chapman and T. Cowling, Mathematical Theory of Non-Uniform
Gases. Cambridge, University Press, 1939.

\bibitem{R} G. R\"opke, Phys. Rev. A {\bf 38}, 3001 (1988).

\bibitem{RR}
G. R\"opke and R. Redmer, Phys. Rev. A {\bf 39}, 907 (1989);\\
R. Redmer, G. R\"opke, F. Morales, and K. Kilimann, Phys. Fluids B {\bf
  2}, 390 (1990);\\
H.Reinholz, R.Redmer and S.Nagel, Phys. Rev. A {\bf 52}, 5368
(1995);
A. Esser and G. R\"opke, Phys. Rev. E  .

\bibitem{sonne}
J.N. Bahcall, M.H. Pinsonneault: Rev. Mod. Phys. {\bf 67}, 781 (1995)

\bibitem{S}
W. Theobald, R. H\"a\ss ner, C. W\"ulker, and R. Sauerbrey,
Phys. Rev. Lett. {\bf 77}, 298 (1996);\\
P. Gibbon, D. Altenbernd, U. Teubner, E. F\"orster, P. Audebert,
J.-P. Geindre, J.-C. Gauthier, and A. Mysyrowicz, Phys. Rev. E {\bf
  55}, R6352 (1997).

\bibitem{mahan}
G. Mahan , {\it Many-Particle Physics} (Plenum, New York, 1981).

\bibitem{rasolt}
M. Rasolt, Phys. Rev. B {\bf 27}, 5653 (1983)
\bibitem{KlimKraeft} Yu.L. Klimontovich and W.D. Kraeft,
    Teplofizika Vyssokich Temperatur (UdSSR) {\bf 12}, 239 (1974). 

\bibitem{Ort}
J. Ortner, F. Schautz, and W. Ebeling, to be published in Phys. Rev. E.

\bibitem{Z} D.N. Zubarev, {\it Nonequilibrium Statistical
    Thermodynamics} (Plenum, New York 1974);\\
D.N. Zubarev, V. Morozov and G. R\"opke, {\it Statistical Mechanics
of Nonequilibrium Processes} (Akademie Verlag, Berlin, 1996, 1997).

\bibitem{KKER}
W. D. Kraeft, D. Kremp, W. Ebeling, and G. R\"opke, {\it Quantum
Statistics of Charged Particle Systems} (Plenum, New York, 1986).

\bibitem{WT}
J.Ward, Phys. Rev. {\bf 78}, 182 (1950). \\
Y.Takahashi, Nuovo Cimento {\bf 6}, 370 (1957). \\
T. Tayoda, Ann. Phys. (N.Y.) {\bf 173}, 226 (1987).

\bibitem{BaymKad}
G. Baym and L. Kadanoff, Phys. Rev. {\bf 124}, 287 (1961). \\
G. Baym, Phys. Rev. {\bf 127}, 1391 (1962).   

\bibitem{wir}
G. R\"opke and A. Wierling, Z. Phys. Chem., in press.\\


\end{thebibliography}
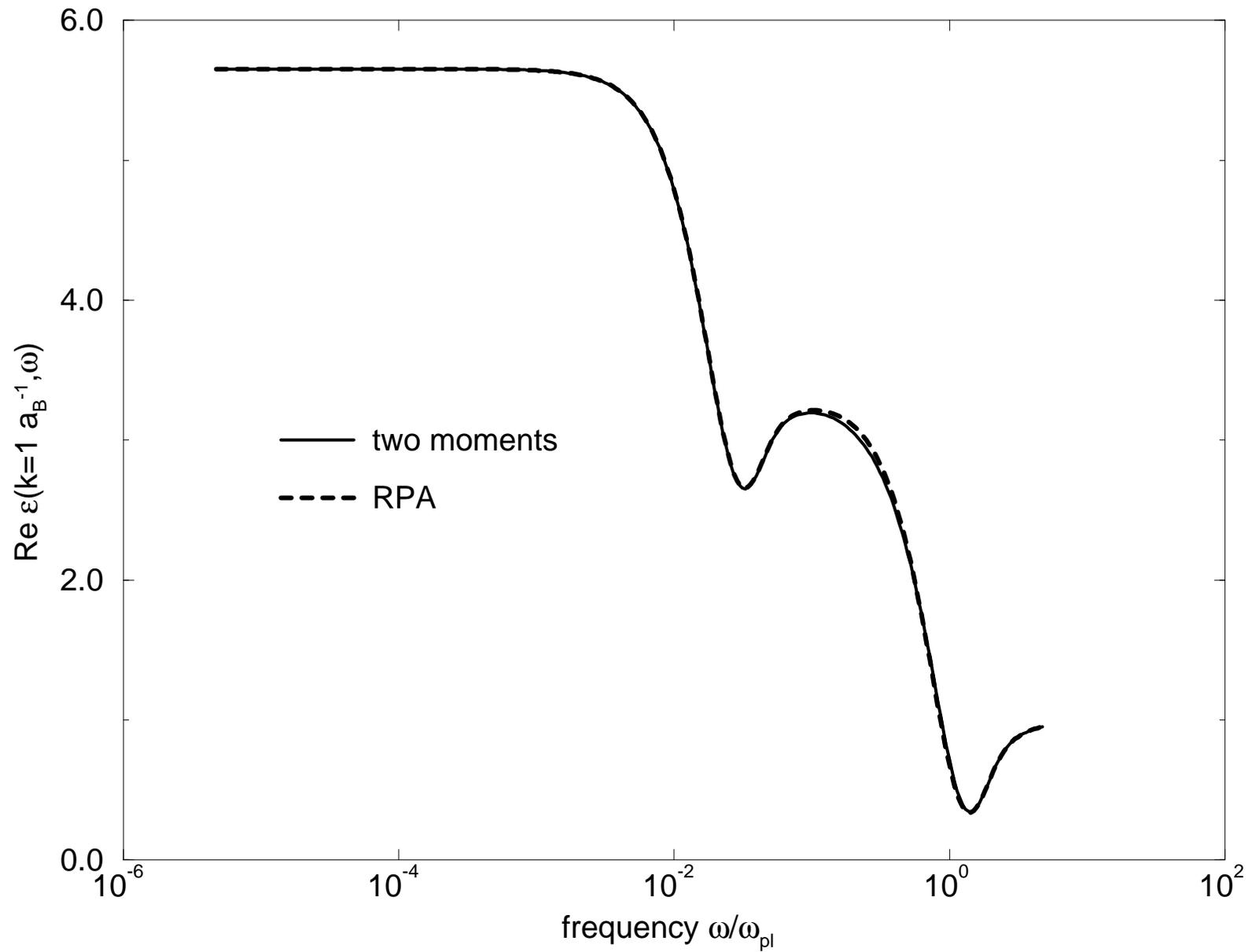
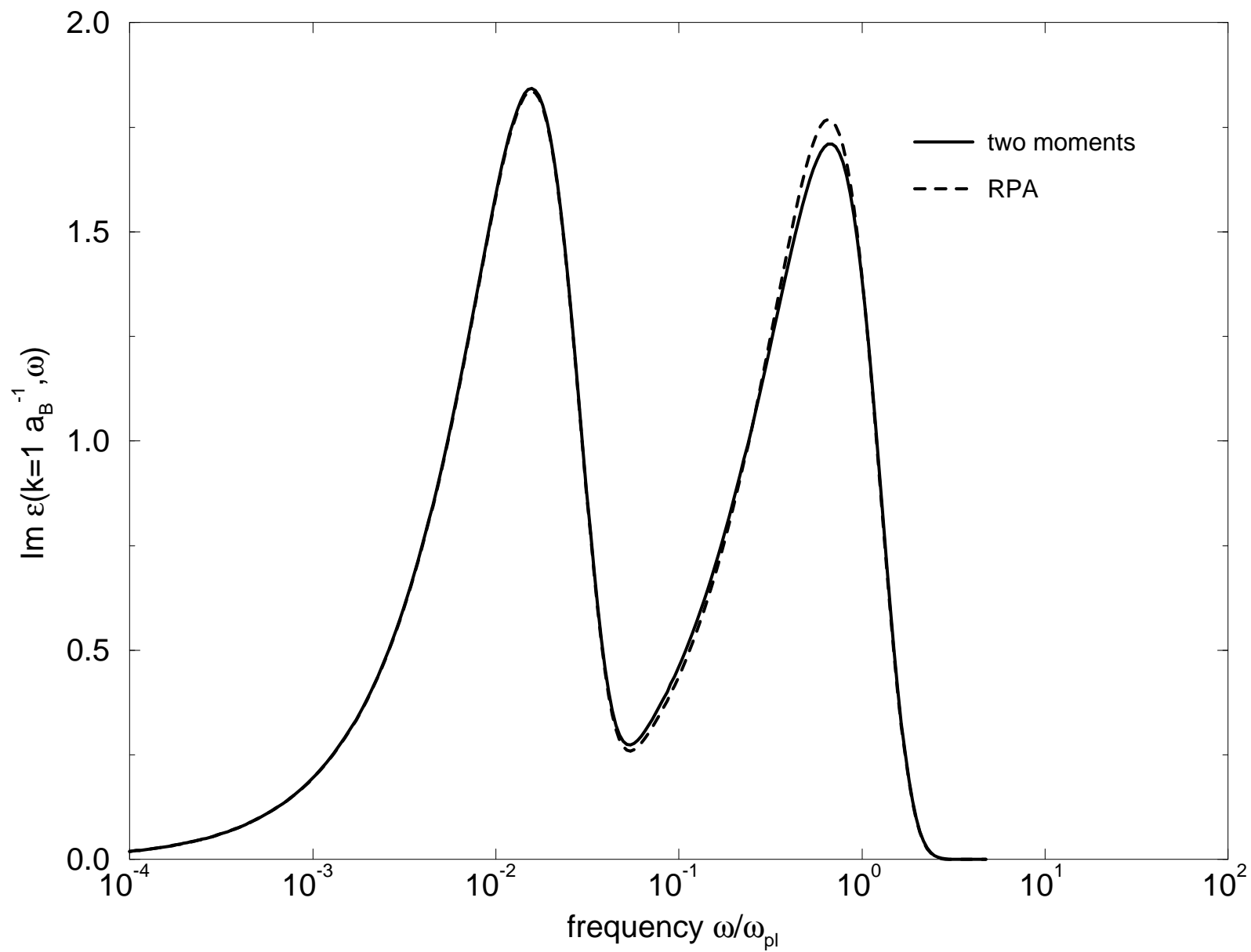
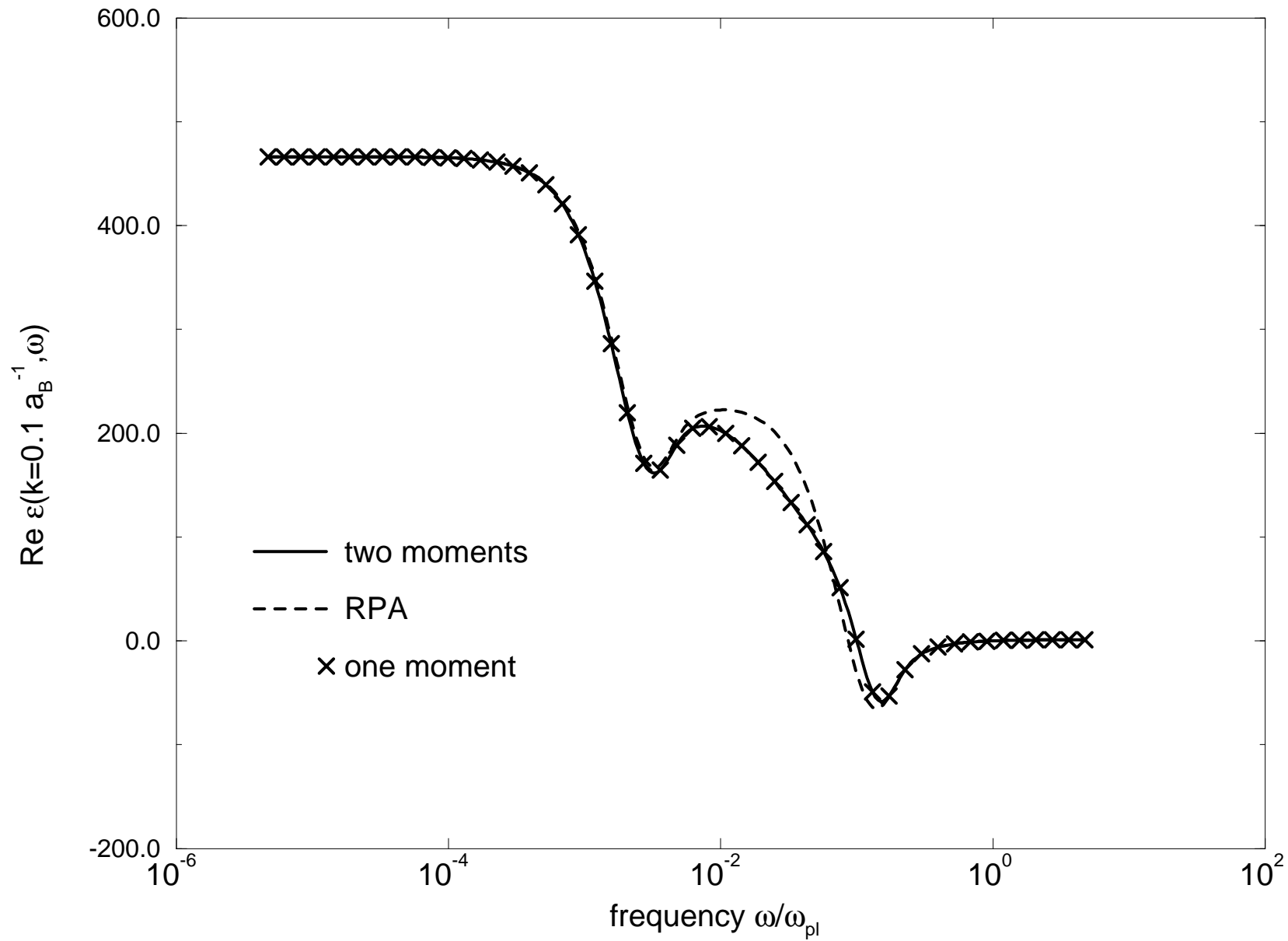
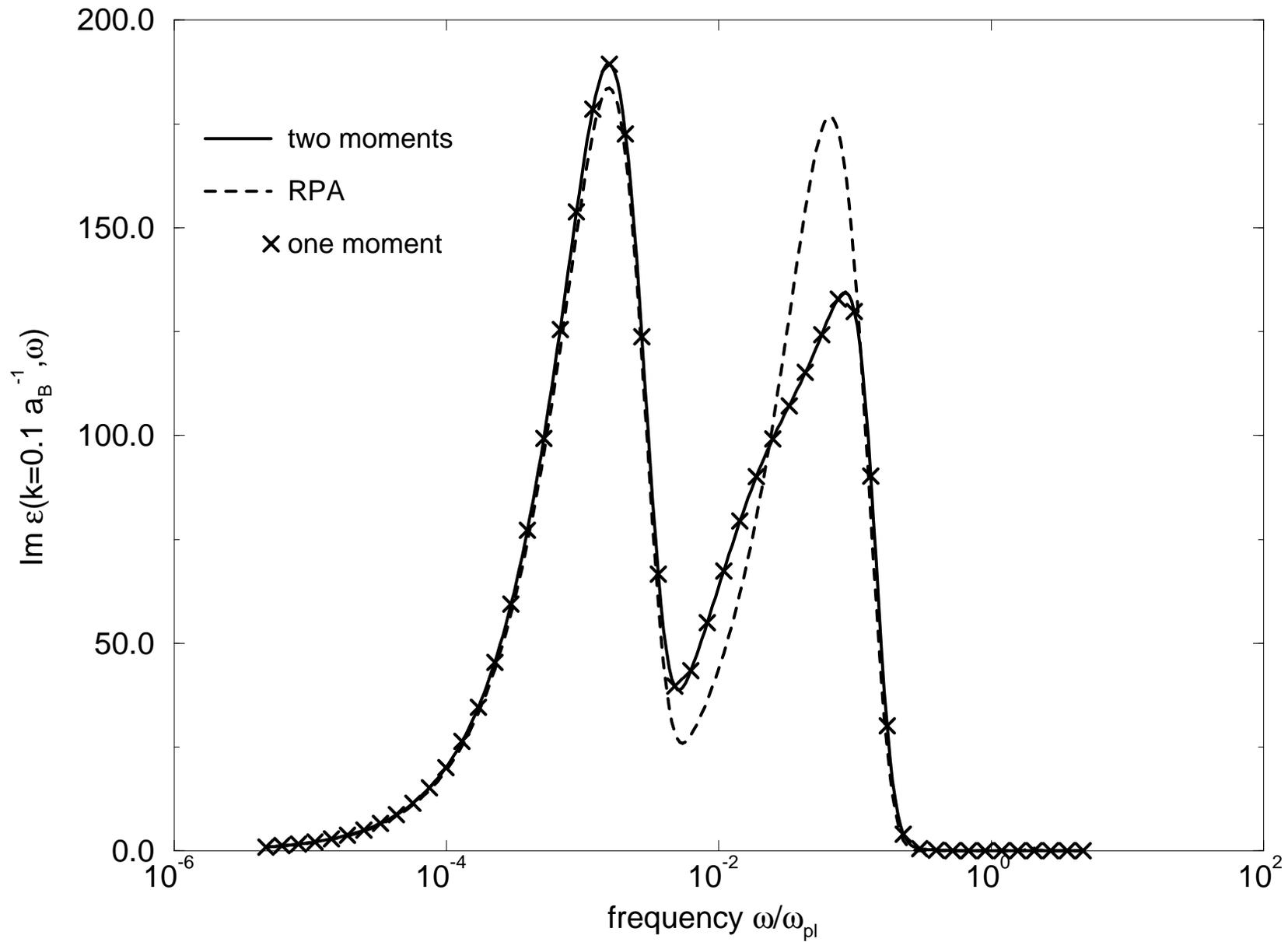
\end{document}